# Fading Model Deviation in The NLOS Communication Channel in Limited Reflection

Zabihollah Hasanshahi , Paeiz Azmi, and Mohammad Khajezadeh

*Abstract*—Statistical models are employed to characterize the clutter in the radar and the reflective signals of the telecommunication receivers. End to this, Rayliegh distribution is the simplest fading models in NLOS channels possessing low-accuracy in the high-resolution radars and distant telecommunication receivers. At present, high accuracy models such as the m-type Nakagami and hybrid GG distributions are utilized in order to model fading. However, despite the Non-Rayliegh models have better precision in the NLOS relative to the Rayliegh models, **the** accuracy of these models decreases **when the radiation angle** in the transmitter and the reflection angle in the receiver are **different**. In this paper, the K distribution function is analytically introduced and deployed to model the fading using practical data. **Although this model was previously introduced to describe the clutter** properties of the radar.

*Index Terms*— NLOS Channel, Non Rayliegh Fading, Limited Reflection

## I. Introduction

The propagation of electromagnetic waves in the wireless channels leads to complex phenomena such as shadows and multipath that in the analysis of communication systems, the precise mathematical description of which is either unrecognized due to the random nature of these phenomena, or very complicated. However, remarkable efforts have been made to statistically model the characteristics of these different effects, which resulted in the determination of a wide range of relatively simple and accurate statistical models, for fading channels extracted under a specific communication scenario in a particular propagation environment.

When the received signal fades away during the transmission, its phase and amplitude fluctuate over time. For coherent modulation, the effects of fading in the phase can greatly reduce the function of the system unless proper actions are taken to compensate these effects in the receiver. Oftentimes, in the analysis of such systems, it is assumed that the coherent ideal demodulation is also employed in the receiver. For non-coherent receivers, phase information is not required in the receiver, therefore, the phase variation arisen from fading have no effect on the receiver's efficiency. Therefore, the performance analysis for different receivers in the fading channels requires the statistical knowledge of the amplitude of the received signals [fading book].

Parameters estimation of the clutter statistical model is an important issue in its modeling, simulation, classification, and identification. The clutter inherently has a random nature; therefore statistical distributions are generally utilized to better determine its characteristics. The K distribution is one of the prevalent models for characterizing the clutter. In estimating the parameters of statistical models based on reception of multi-path signals in telecommunications, the mentioned models are also acceptable, but in order to explain different aspects of the problem, in this study, the signals received from the transmitter in multipath mode from omnidirectional and directional antenna (from the main, the secondary and the reflective lobes) and the results are analyzed and the deviation of the fading model in the mode of receiving multi-path signals from the reflective lobe employing the GOF criteria is compared. In the second section, various Non-Rayliegh models deployed in the modeling of fading and shadow phenomena are evaluated, and then in the third section, the Kullback-Liebler, Kolmogrof-Smirnof and the Anderson-Darling criterions are introduced and model evaluation indicators are analyzed based on practical data. In Section 4, signals from fixed mobile stations in omnidirectional and limited reflection modes are received by employing directional antennas and by using different channel estimation methods and information receiving, the fading model of receiving channel is obtained, afterward, by utilizing the mentioned criteria in the section 3, the results are assessed. Finally, in the section 5, the results are summarized in accordance with various charts like KS-Test.

## II. Non-Rayliegh Models for Random Channels

In order to design a multi-path experiment, it is essential to take a mathematical model into account [1]. In this model, it is assumed that the propagation environment operates as a linear filter so that if $\text{Re}\{s(t)e^{j\omega_0 t}\}, t \in (-\infty, \infty)$ in which s(t) is the baseband mixed-signal and $\omega_0$ is the carrier frequency it will be sent. The signal $\text{Re}\{\rho(t)e^{j\omega_0 t}\}, t \in (-\infty, \infty)$ is obtained, while:

$$\rho(t) = \sum_{0}^{\infty} a_k s(t-t_k)e^{i\theta} + n(t) \qquad (1)$$

Therefore, the environment can be described utilizing the baseband accumulative noises as well as the following statistical variables:

$$\{a_1, a_2, a_3, \ldots t_1, t_2, t_3 \ldots \theta_1, \theta_2, \theta_3, \ldots\} = \{a_k, t_k, \theta_k\}_0^{\infty}$$

It can be said that n(t) is well-compatibled at least in the intended frequencies with the Gaussian model at least at the frequencies considered, and therefore there is no interest in the evaluation of its characteristics in this study. Also, since it is obtained from independent paths and its sensitivity to the properties of each path is very large, it is possible to consider the uniform model in the interval $(0, 2\pi]$. In other studies, the variable $t_k$ has been studied and the Poisson model is considered for it [2], [3]. In this paper, it is tried to investigate the variable $a_k$, which is the amplitude of the fading paths, in a different way from what was accomplished in previous studies. Urban radio channels are inherently part of multipath channels due to the reflection, reflection, and dispersion or scattering of the buildings in the wave. The path strength in a specific delay is the result of summation of all paths that have the same distance, up to the receiver. The strength and the phase of each path, as well as the number of these paths, determines the signal strength in that particular delay. It is totally accepted that the Rayleigh and Rician distributions describe the power of paths of less than a few hundred wavelengths, and the log-normal distribution is appropriate for environments with more than this distance. Also, in many paper, combinational distributions have been deployed to describe the power of each path, the general form of which is as follows:

$$p(x) = \int_{\Sigma} f(x,\sigma) dQ(\sigma) \quad (2)$$

Where $f(x,\sigma)$ is the distribution function taken for the local environment around the receiver, which can include n parameters and $Q(\sigma)$ is the distribution of $\sigma$. For instance, if the local area is considered as Rayleigh and the transmitter path is model as log-normal (due to the remoteness of the receiver's local environment) the general distribution in relation (2) is obtained as follows:

$$p(x) = \int_0^\infty \frac{x}{\sigma^2} \exp(-x^2/2\sigma^2) \frac{1}{\sqrt{2\pi}\sigma\lambda} \cdot \exp\left(\frac{(\log\sigma - \mu)}{2\lambda^2}\right) d\sigma \quad (3)$$

*A. K Fading Model*

Due to the complexity of the above equation, the closed-form expression of the log-normal distribution is approximated by the gamma distribution, and the result of (3), is the distribution function k [4] as follows:

$$f_k(t) = \frac{2c}{\Gamma(\upsilon)}\left(\frac{ct}{2}\right)^\upsilon K_{\upsilon-1}(ct) \quad (4)$$

where $\Gamma(0)$ is the gamma function, $K_\upsilon$ is the type two $\upsilon$ order modified Bessel function. The parameters c and v are called shape and scale, respectively. It could be proved that, for $\upsilon > 20$ the distribution K tends to the Rayleigh function [5].

*B. F Fading Model*

Use one space after periods and colons. Hyphenate complex modifiers: "zero-field-cooled magnetization." Avoid dangling participles, such as, "Using (1), the potential was calculated." [It is not clear who or what used (1).] Write instead, "The potential was calculated by using (1)," or "Using (1), we calculated the potential."
In a composite fading channel, it is possible to accord the multipath channel characteristics with the Nakagami model. Meanwhile, it is assumed that the RMS signal power is a random variable that variates by the influence of shadow noise. In this case, the signal envelope can be given as $R^2 = \sum_{n=1}^{m} A^2 X_n^2 + A^2 Y_n^2$ while **m** is indicative of path numbers, $X_n$ and $Y_n$ are the and quadrature variables of the signal which are independent and Gaussian. If **A** is considered in the form of the inverse Nakagami distribution with the in-shape parameter $m_s$ and the scale parameter $1/\Omega_s$, then it can be concluded that [6]:

$$f_A(x) = \frac{2m_s^{m_s}}{\Gamma(m_s)\alpha^{2m_s+1}} \exp(-\frac{m_s}{\alpha^2}) \quad (5)$$

Therefore, the PDF envelope of the signal can be obtained by averaging from Nakagami distribution on the random variable of the RMS signal power meaning $f_R(r) = \int_0^\infty f_{R|A}(r|\alpha) f_A(\alpha) d\alpha$. Considering the taken model of the signal, it can be expressed:

$$f_{R|A}(r|\alpha) = \frac{2m^m}{\Gamma(m)\alpha^{2m}} \exp(-\frac{mr^2}{\alpha^2\Omega}) \quad (6)$$

In which m is the fading intensity parameter, and $\Omega = E[r^2]$ is the average RMS power. By changing the variable and integration on the equation 6, it can be deduced:

$$f_R(r) = \frac{2m^m (m_s\Omega)^{m_s} r^{2m-1}}{B(m,m_s)(mr^2 + m_s\Omega)^{m+m_s}} \quad (7)$$

Where $B(0,0)$ is the Beta function [7]. The equation 7 is called the distribution K.

*C. KG Fading Model*

In the discussion of distribution F, it was said that the effects of shadow noise on average RMS power are modeled in the form of inverse Nakagami distribution. However, if all the discussions of the F distribution section remain constant and transform only the average power model of the signal into gamma distribution, the obtained model is called Gamma-Gamma or Generalized-K [8]. Initially, a generalized k model was used to model the propagation of radar and signal ecosystems in a sonar [9] and [10], however, nowadays it has been utilizing in order to model wireless telecommunication fading channels. With some mathematical simplifications, the formula for the generalized k model can be expressed by:

$$f_X(X) = \frac{4m^{(\beta+1)/2} X^\beta}{\Gamma(m)\Gamma(k)\Omega^{(\beta+1)/2}} K_\alpha \left[2\left(\frac{m}{\Omega}\right)^{1/2} X\right] \quad (8)$$

In which $\alpha = k - m$, and $\alpha = k + m - 1$. while k, m are the shape parameters. $K_\alpha(0)$ is the modified Bessel function by the degree α. Also $\Gamma(0)$ is the gamma function and $\Omega = E\langle X^2 \rangle / k$ is the average power. In fact, the KG distribution has two parameters that by variating them, other distributions can be estimated, namely if in the above equation $k \to \infty$ then the formula (8) estimates Nakagami-m distribution. For m=1 it turns to K distribution. Also if $m \to \infty$ and $k \to \infty$ it tends to Gaussian model.

Due to the high complexity, it is usually not possible to employ this model to derive other formulas, for instance, in [11], it is tried to obtain an appropriate approximation of this distribution utilizing gamma distribution.

## III. GOF TEST

The point of GOF is the methods indicating that how much the sampled data are accorded with the assumed distribution function. In the Null hypothesis (formal test frame) $H_0$ is indicative of a state in which the random variable X follows the probability function F(x) (for instance normal function or Weibull). The GOF methods normally introduce the criterion by which the degree of data matching is measured in $H_0$ on the assumed distribution function. In many cases, the application of GOF is limited to the alternative hypothesis or the complex H1 and usually conveys a few information and simply states that $H_0$ is not correct. It is obvious that the concentration is on the measurement of the data matching degree with $H_0$ and it is usually expected that H0 hypothesis is demonstrated in the test. In this paper, three independent GOF methods are assessed. The first and second methods are the subset of continuous methods based on measurement of experimental function PDF, also known as EDF [12] and the third is based on the Kernel method and utilizing the Kullback-Leibler meter which is a merely comparative method [13] explained briefly in the following section.

### A. Statistical Characteristics Based on EDF

Assuming there is n sample of random variable denoted by $X_{(1)},....,X_{(n)}$ and respected as $X_{(1)} < X_{(2)} < .... < X_{(n)}$. Hence the EDF function can be expressed as:

$$F_n(X) = \frac{\text{Number of Observeations} \leq X}{n} \quad (9)$$

The statistics measuring the difference between $F_n(X)$ and $F(X)$ are called EDF statistics. All of this statistics operate based on vertical difference between $F_n(X)$ and $F(X)$. and are divided into two major categories that their difference has roots in the meter type deployed in the test. In the first type, one of supremum-based meters and in the second type a quadratic-based meters have been used. Kolmogrof-Smirnof test is chosen from the first type and Anderson-Darling is selected from the second type.

The statistical characteristic of Kolmogrof-Smirnof (10) and Anderson-Darling are defined as:

$$D = \sup_X |F_n(X) - F(X)| \quad (10)$$

$$A = n\int_{-\infty}^{\infty} \{F_n(X) - F(X)\}^2 \psi(x) dF(X) \quad (11)$$

In which $\psi(x) = \left[\{F(X)\}\{1 - F(X)\}\right]^{-1}$.

### B. α Levels

It should be noted that the more $F_n(X)$ is closer to the proposed F(x) distribution the less value is the test result and this probability that the data will really follow the F(x) is higher. However, in GOF tests, there is actually no definitive idea of whether or not the data would follow from F(x) and only the α level is considered for the correctness of $H_0$. For the explanation of α level, let's assuming that the result of the test T for the samples $X_{(1)},....,X_{(n)}$ is equal to the t. α levels are defined as $p = P(T > t | H_0)$. In other words, this levels explain that if p=α, this means that for the α probability supposing $H_0$ is true the value of the test T becomes more than the value obtained from the above tests for its indicative of there is a slight probability for correctness of $H_0$ and a higher value from the test T is obtained for the result. For instance, α levels for the mentioned test in the previous section is given below:

TABLE I
α-levels for the tests A and D [12]

| Test Type | α-levels | | |
|---|---|---|---|
| | 0.25 | 0.05 | 0.001 |
| D | 0.828 | 1.224 | 1.859 |
| A | 1.248 | 2.492 | 6.000 |

### C. Kullback-Liebler Divergence

Kullback-Liebler divergence or briefly (KLD) [14] is a measurement criterion of information theory measuring the difference between two given distributions. If the random variable z is given and also p(z) and q(z) are two distinct distributions, then the KLD is defined as follows:

$$D(P,Q) = \int p(z) \log \frac{p(z)}{q(z)} dz \quad (12)$$

The value of KLD is always non-negative and is zero only for the state in which p=q. KLD can also be defined as expectation value of the below function:

$$\lambda(z) = \log \frac{p(z)}{q(z)}$$

Therefore, the more $\lambda(z)$ is closer to the zero (or p and q closer to each other) the less KLD will be. As it can be seen, the KLD is an analytical statistical characteristic and it requires two closed-form PDF (not experimental). Hence, for the comparison of different distributions with the obtained data, a closed-form PDF should be taken into account which in this study, mentioned PDF is attained from the Kernel estimation.

*D. Kernel Estimation*

If there is *iid* N sample in the form of $X_{(1)},....,X_{(n)}$ from a probability distribution f(x), the Kernel estimation of which is defined as:

$$\hat{f}(x) = \frac{1}{nh}\sum_{i=1}^{\infty} K\left(\frac{x-X_i}{h}\right) \quad (13)$$

Where K is the probability distribution function with the condition $\int_{-\infty}^{\infty} K(x)dX = 1$ and h is the width of the window.

As it is evident, the Kernel estimation accuracy depends on the h and K. for instance if h is selected small the expectation value reduces but the estimation variance increases instead and conversely will happen if h decreases. Also, the optimal value of h and K can be estimated by the definition of MISE criterion as follows:

$$MISE(\hat{f}) = E\int_{-\infty}^{\infty} \left\{\hat{f}(X) - f(x)\right\}^2 dX \quad (14)$$

Using this criterion optimal value for h is obtained as:

$$h_{opt} = k_2^{2/5} \left\{\int K(t)^2\right\}^{1/5} \left\{\int f''(t)^2 dt\right\}^{-1/5} n^{1/5} \quad (15)$$

In which $k_2 = \int t^2 f(t)dt$ is optimal estimation of h to the dependent probability distribution function that is unfortunately unavailable, however, considering above formula, it can be concluded that increase in samples results in the slow decrease rate for the optimal h. Furthermore, due to $k_2 = \int f''(t)^2 dt$ measures fluctuation value of the distribution function, it can be deduced that lower value of the h for the functions having wide range fluctuations is appropriate.

If the optimal value of the is returned to the equation (14), MISE is only the function K which by minimizing it, Epanechnikov Kernel for the K is acquired, defined as:

$$K_e(t) = \begin{cases} \frac{3}{4\sqrt{5}}\left(1 - \frac{1}{5}t^2\right) & -\sqrt{5} \le t \le \sqrt{5} \\ 0 & \text{Otherwise} \end{cases} \quad (14)$$

IV. FADING MODEL DEVIATION BASED ON CHANNEL ESTIMATION

In the propagation environment of mobile signal in each time interval, the channel response is a random and unrecognized function. Also in the TDMA-based communication systems information bits have been transmitted and received in the form of Bursts, therefor every channel estimation algorithm such as blind or non-blind should abled to estimate a trustable estimation of the channel response from few received samples. Blind methods are not utilized in the operational GSM systems [15] due to practically it is in need of thousands (or more) of samples for a valid estimation. GSM systems employs the frequency range 880-915 MHz for the uplink and 925-960 MHz for the downlink. These bandwidths in the next phase will be divided into 200 kHz bands and by the assistant of the ARFCN, numbers become distinct. Afterwards, this channels divide into the octet TDMA frames by the width of 576 μs. Each time slut can be deployed by the transmitter or receiver for transmitting or receiving Bursts. As it is illustrated in the fig.1. a SCH Burst is composed of 64 train bit and some information bits. For further information about GSM networks, it should be referred to [16].

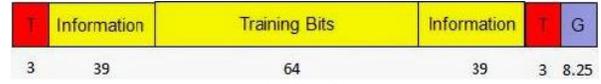

Fig.1 SCH Bursts

As it is mentioned above, the Bursts of GSM include predetermined bits, generally utilized to synchronize transmitter and receiver. For instance, FCCH blocks transmit zero bits sequentially which result in sinusoidal pulse in the baseband with the frequency 67 kHz (fig. .2.).

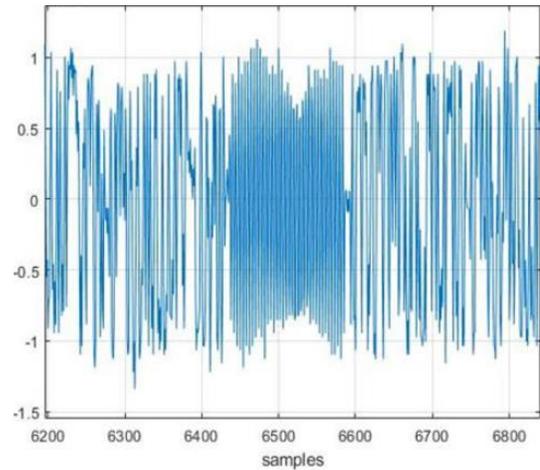

Fig. 2. Real part of the baseband signal which includes FCCH

After transmitting FCCH block, transmission of SCH Burst will be in turn in which 64 train bit, exists as follows:

SCHbits= [1 0 1 1 1 0 0 1 0 1 1 0 0 0 1 0 0 0 0 0 0 1 0 0 0 0 0 0 1 1 1 1 0 0 1 0 1 1 0 1 0 1 0 0 0 1 0 1 0 1 1 1 0 1 1 0 0 0 0 1 1 0 1 1]

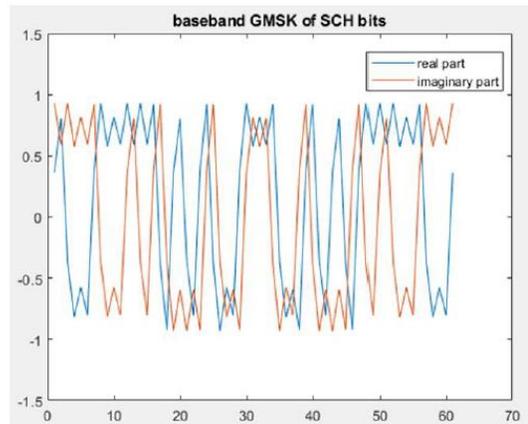

Fig.3. Transmitted baseband signal equaling to the train bits in the GMSK modulation

The transmitting signals as the above bits employed in GMSK modulation in the GSM systems are indicated in fig. 3. Also, serial Bursts of the GSM signals are depicted in fig. 5. It should be notified that the fig. 2. and fig. 5. are sampled in the Matlab software using USRP 210 and the rate equal to the 1625k/6 (270.83 Ksample/s according to the GSM standard).

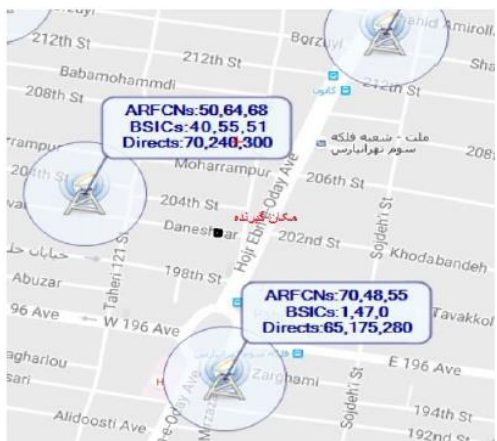
Fig.4 Location of the transmitter and receiver

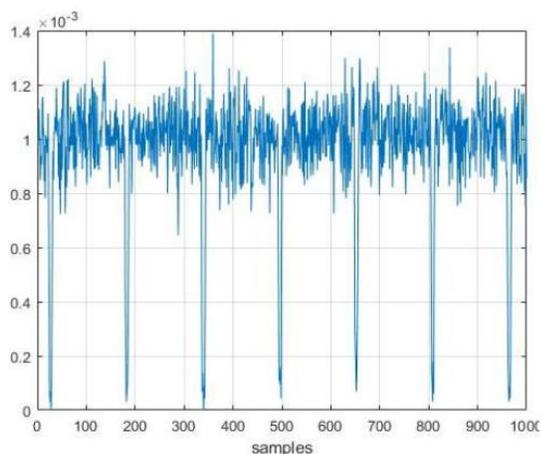
Fig. 5. Serial bursts of the GSM network's controlling channel

As it is evident from fig. 3. the transmitting signals are equal to specified train bits represented by $s_r(n)$. Also using Matlab Software by decoding the received GSM signals and tracking the SCH bits after observation of FCCH, received signals equaling to the train bits can be obtained which are represented by…. . Therefore, the channel can be estimated for the frequency interval in which transmitting signal is not zero.

$$s_r(n) = h(n) * s_t(n)$$
$$h(n) = F^{-1}\left\{S_r(e^{j\omega}) / S_t(e^{j\omega})\right\} \quad (14)$$

Since the signal sampling is accomplished by the rate 270.83 Ks/s, each channel tap is equal to 1/270.83K or 3.7 µs in the channel estimation. The location of experiments along with the location of BTSs are illustrated in fig. 4.

According to the fig. 4. Each spot is illustrative of 3 BTSs with the specified ARFCN numbers in the first line. Also in the third line, the numbers related to the antenna's angle is given as direction. These numbers are write in degree and relative to the north of the map. Two BTSs with the ARFCN 50 and 48 are evaluated. The receiver is located in main lobe relative to the BTS number 50 and also is located in the back lobe relative to the BTS number 48. The central frequency of the transmitter number 50 is 945 MHz and is 944.6 MHz for the transmitter number 48. In both modes, signal is recorded and the channel is estimated approximately 3000 times. Fig. 6. depicts foe ensemble from the random procedure |h(n)| with 14 taps.

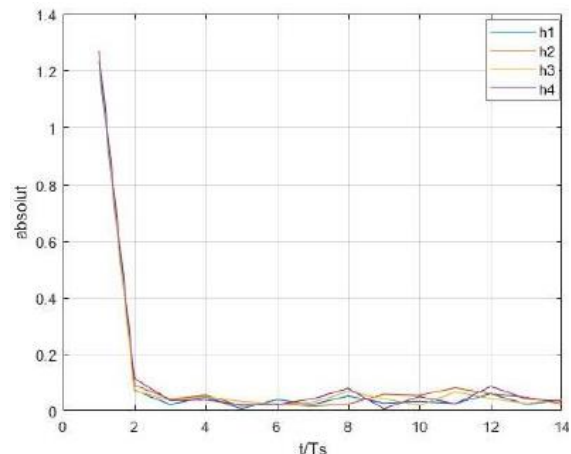
Fig. 6. For random samples from random procedure of the channel response measurement

As it can be seen from the fig. 6. the channel is estimated by 14 taps. In the following, all of the taps are compared with each other by the statistical tests. Initially, the signal of the BTS number 50 that is received in the main lobe of the antenna is analyzed. Fig.7. indicates the data histogram related to the forth tap of the channel estimation. Also, estimated PDFs by the MLE method along with the kernel estimation of the data are given.

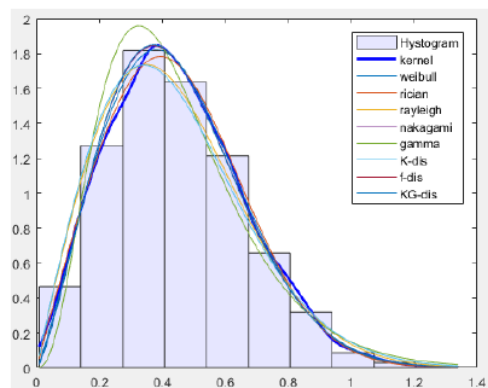
Fig. 7. The histogram of the forth tap of the BTS channel number 50

To compare different PDFs with the introduced statistical method in the previous section, firstly, the parameters of each PDF are estimated by the MLE technique and then α-levels are obtained for each PDF employing acquired PDF and the existing data.

If the α-level for the result of each test is more than 5% the null hypothesis is confirmed (which means the considered PDF is matched with the data).

TABLE II
statistical tests for the forth tap of the BTS channel 50

| PDF | Null hypothesis | α-level A-D | α-level K-S | -20log KLD |
|---|---|---|---|---|
| Weibull | confirmed | 0.95 | 0.82 | 28.8 |
| Rician | confirmed | 0.65 | 0.83 | 29.6 |
| Rayliegh | Rejected | 1.7e-5 | 1.2e-5 | 20.4 |
| Nakagami | confirmed | 0.66 | 0.44 | 26 |
| Gamma | Rejected | 5.06e-5 | 3.2e-5 | 15 |
| K | Rejected | - | 3.2e-7 | 18 |
| F | confirmed | - | 0.34 | 24 |
| KG | confirmed | - | 0.33 | 22 |

As it is observable from the result of the table 2, the forth tap of the BTS number 50 confirms the Rician distribution with the strongest α-level and the Nakagami distribution function also appears to be appropriate for this data but the channel estimation includes 14 taps. Fig. 8. illustrates the α-levels for comparing in the K-S test for other taps.

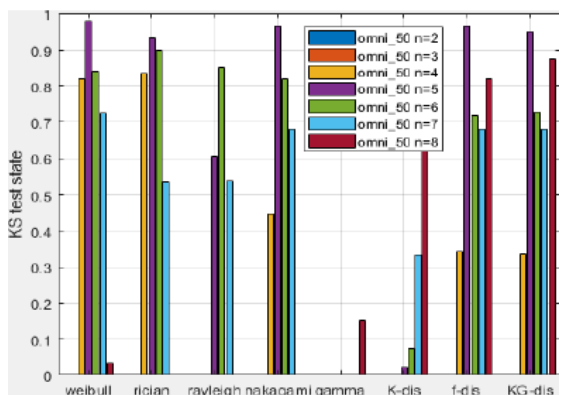
Fig. 8. Delay comparison of the BTS channel number 50

It is observed that the delays number 2 and 3 are not compatible with any of the existing PDFs. The delay number 4 confirms Rician and Weibull distributions and the delay number 5 validates to many of other distributions and the delay number 8 only authenticates the F, K, KG PDFs. The second experiment is conducted similar to the BTS number 48 with this assumption that since we are located in the back lobe of the antenna and the signal receives generally after reflection from obstacles, it should have a different nature from previous experiment.

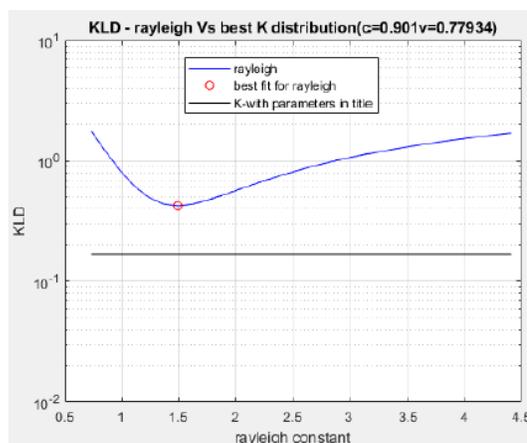
Fig. 9. The comparison between rayliegh and K distributions for the tenth delay of the BTS number 48

In the fig. 9. matching of the Rayliegh and K distributions for the $10^{th}$ tap of channel estimation for the BTS number 48 are compared. The horizontal axis indicates the value of statistical characteristic KLD for the resulted PDF from the same Rayliegh constant. It can be seen that the best accordance with the data can be obtained for the Rayliegh constant equal to 1.5. However, the black line of the statistical characteristic KLD for the distribution K with ν= 0.7793, c= 0.901 has a less KLD and, hence, better accordance. Therefore, this channel can be taken as deviated channel from the Rayliegh into account. Table 2 presents the statistical tests related to the $10^{th}$ tap of the channel. Similar to that of previous section, the α-level more than %5 is considered as confirmation of the distribution accordance.

TABLE III
Statistical tests for the forth tap of the BTS channel 48

| PDF | Null hypothesis | α-level A-D | α-level K-S | -20log KLD |
|---|---|---|---|---|
| Weibull | Rejected | 2.45e-7 | 7.55e-8 | 14 |
| Rician | Rejected | 2.45e-7 | 0 | 1.2 |
| Rayliegh | Rejected | 2.45e-7 | 0 | 1.2 |
| Nakagami | Rejected | 2.45e-7 | 0 | 9.1 |
| Gamma | Rejected | 2.45e-7 | 6.6e-9 | 15.2 |
| K | Rejected | - | 1.5e-8 | 15.5 |
| F | confirmed | - | 0.11 | 20 |

Therefore, only the distribution F is compatible with data. To compare different taps of the channel, bar chart is shown in fig. 10. It is evident that only the distribution F is compatible with the data, by which the results are more meaningfully prosperous than the other distributions.

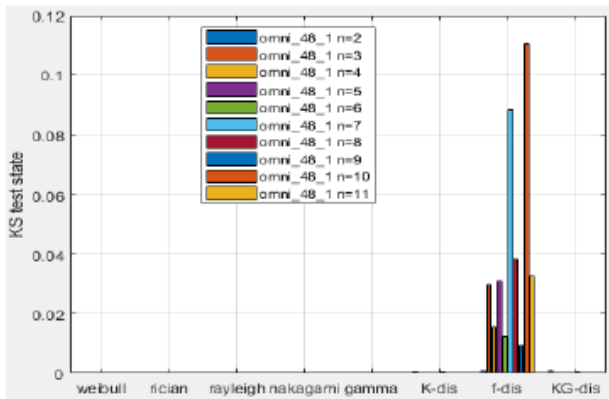

Fig. 10. The comparison between different delays of the BTS 48

## V. Conclusion

In this paper, the problem of the fading model estimation in the NLOS channels for the limited refection mode utilizing directional antennas from the back lobe is analyzed using the accumulated urban channel estimation and the train bits and SCH Bursts of the GSM signals are used for the urban channel estimation. Two type of channels are investigated that the major difference of which is in the angle of the antenna relative to the receiver so that at first test the receiver was located in the main lobe of transmitter without reflection limitation and in the second test was located in the back lobe with reflection limitation. The results of statistical tests accomplished by the variant GOF methods indicates that the first channel can be modeled to acceptable extent by the classic methods stated for the urban fading channels such as Rician, Nakagami, Rayliegh and etc. However, in the back lobe due to the variation of the channel conditions and decreasing in the NLOS paths in compare with the first state and also weakening the in-line-of-sight paths there is no possibility to estimate channels with the mentioned methods, therefor, among all the distributions in this study, the distribution F observed to have better performance.

REFERENCES:

[1] Turin, G.L., W.S. Jewell, and T.L. Johnston, *Simulation of urban vehicle-monitoring systems.* IEEE Transactions on Vehicular Technology, 1972. **21**(1): p. 9-16.
[2] Suzuki, H., *A statistical model for urban radio propogation.* IEEE Transactions on communications, 1977. **25**(7): p. 673-680.
[3] Ganesh, R. and K. Pahlavan. *On the modeling of fading multipath indoor radio channels.* in *Global Telecommunications Conference and Exhibition'Communications Technology for the 1990s and Beyond'(GLOBECOM), 1989. IEEE*. 198 .⁹ IEEE.
[4] Abdi, A. and M. Kaveh, *K distribution: An appropriate substitute for Rayleigh-lognormal distribution in fading-shadowing wireless channels.* Electronics Letters, 1998. **34**(9): p. 851-852.
[5] Weinberg, G.V., *Error bounds on the Rayleigh approximation of the K-distribution.* IET Signal Processing, 2016. **10**(3): p. 284-290.
[6] Yoo, S.K., et al., *The Fisher–Snedecor $\mathcal {F} $ Distribution: A Simple and Accurate Composite Fading Model.* IEEE Communications Letters, 2017. **21**(7): p. 1661-1664.
[7] Gradshteyn, I.S. and I.M. Ryzhik, *Table of Integrals, Series, and Products. Edited by A. Jeffrey and D. Zwillinger.* 2007, Academic Press, New York.
[8] Bithas, P.S., et al., *On the performance analysis of digital communications over generalized-K fading channels.* IEEE Communications Letters, 2006. **10**(5): p. 353-355.
[9] Lewinski, D., *Nonstationary probabilistic target and clutter scattering models.* IEEE transactions on antennas and propagation, 1983. **31**(3): p. 490-498.
[10] Gu, M. and D.A. Abraham, *Using McDaniel's model to represent non-Rayleigh reverberation.* IEEE Journal of Oceanic Engineering, 2001. **26**(3): p. 348-357.
[11] Al-Ahmadi, S. and H. Yanikomeroglu, *On the approximation of the generalized-K distribution by a gamma distribution for modeling composite fading channels.* IEEE Transactions on Wireless Communications, 2010. **9**(2.(
[12] Stephens, M., *Goodness-of-fit techniques.* Series Statistics: Textbooks and Monographs, 1986. **68**.
[13] Weinberg, G.V. and V.G. Glenny, *Optimal Rayleigh Approximation of the K-Distribution via the Kullback–Leibler Divergence.* IEEE signal processing letters, 2016. **23**(8): p. 1067-1070.
[14] Kullback, S. and R.A. Leibler, *On information and sufficiency.* The annals of mathematical statistics, 1951. **22**(1): p. 79-86.
[15] Tong, L., G. Xu, and T. Kailath, *Blind identification and equalization based on second-order statistics: A time domain approach.* IEEE Transactions on information Theory, 1994. **40**(2): p. 340-349.
[16] ETSI, G., *03.20:" Digital cellular telecommunications system (Phase 2* .(+Security related network functions, 1992.